\documentclass[fleqn,twoside]{article}
\usepackage{espcrc2}

\newcommand{\be}{\begin{equation}}
\newcommand{\ee}{\end{equation}}
\newcommand{\bea}{\begin{eqnarray}}
\newcommand{\eea}{\end{eqnarray}}

\usepackage{graphicx}
\usepackage[figuresright]{rotating}

\title{The Kaon B-parameter from Two-Flavour Dynamical Domain Wall Fermions.}

\author{C. Dawson
\address{RIKEN-BNL Research Center,Bldg 510a, Upton, NY 11973-5000}
[RBC Collaboration]
\thanks{We thank RIKEN, Brookhaven National Laboratory and the U.S.\ Department
of Energy for providing the facilities essential for the completion of
this work.}}

\begin{document}

\begin{abstract}
We report on the calculation of the kaon B-parameter using two dynamical
flavours of domain wall fermions. Our analysis is based on three ensembles of
configurations, each consisting of about 5,000 HMC trajectories, with a
lattice spacing of approximately 1.7 GeV for $16^3\times 32$ lattices;
dynamical quark masses range from approximately the strange quark mass to half
of that. Both degenerate and non-degenerate quark masses are used for the
kaons.
\end{abstract}
\maketitle

\section{INTRODUCTION}

The kaon B-parameter, $B_K$, is one of the simplest quantities that may be
calculated on the lattice which may be used to directly constrain the
unitarity triangle.  However, its calculation on the lattice is greatly
complicated when either flavour or chiral symmetry is broken by the lattice
action used.  When using the domain wall fermion lattice action, flavour
symmetry is exact, and the explicit breaking of chiral symmetry is small
enough to be neglected. Because of this it is natural to use domain wall
fermions for the calculation of $B_K$, and, indeed, this has been done to
great success in the quenched approximation \cite{cd:jun}.  Here we report on
the calculation of $B_K$ as part of the first large-scale dynamical
(two-flavour) domain wall fermion simulations.  As well as studying the
effects of un-quenching, we report on the effects of including non-degenerate
masses ($m_s \neq m_u = m_d$) in both the lattice calculation and the
extrapolation to the physical point.

\section{DYNAMICAL ENSEMBLES}

The results of this work are based on three ensembles of two-flavour dynamical
domain wall fermion configurations.  Each ensemble utilises a lattice of size
$16^3\times 32$, a domain wall height of 1.8, a fifth dimension of extent 12,
and consists of 94 configurations with the separation of 50 HMC trajectories
between each configuration.  The bare quark masses of these ensembles are
0.02, 0.03 and 0.04 respectively, the latter of which roughly corresponds to
the strange quark mass. Details of lattice generation, and further physical
results can be found in \cite{cd:dyp,cd:taku04,Dawson:2003ph}; here we will
touch on only a couple quantities of relevance to the extraction of the kaon
B-parameter: the residual mass and the lattice spacing. All quantities are
quoted with a jackknife error (although, where relevant, the average
correlation matrix was used).   

The residual mass, $am_{\rm res}$, is the additive shift in the mass due to
the residual chiral symmetry breaking of domain wall fermions. We calculate
this from the breaking term in the ward identity \cite{Blum:2000kn},
extracting our final value of $am_{\rm res} = 0.001372(49)$ by linearly
extrapolating to zero dynamical mass.  The lattice spacing we use is fixed by
comparing a dynamical extrapolation of the vector meson to the physical point
to the experimental mass of the rho meson. This gives value of $a^{-1} =
1.691(53) {\rm GeV}$.  It is interesting to note, however, that should we fix
the lattice spacing from the Sommer parameter, the result is consistent
\cite{cd:koichi}.
\vspace{-0.1cm}
\section{$B_K$}
In the continuum, $B_K$ is defined as
\begin{eqnarray}
B_K &=& 
\frac{\langle \bar{K^0}|O_{LL}|K^0\rangle}
{\frac{8}{3}\,f_K^2\,M_K^2}.
\\
&=&
\frac{\langle \bar{K^0}|
 O_{LL} |K^0\rangle}
{\frac{8}{3}\,\langle \bar{K^0}|A_4|0\rangle\langle 0|A_4|K^0\rangle},
\label{eq:bk}
\end{eqnarray}
where $A_4$ is the time component of the axial current, and 
\be O_{LL} =
\bar{s}\gamma_\mu(1-\gamma_5)d\,\bar{s}\gamma_\mu(1-\gamma_5)d \, .  
\ee
When using a lattice action which breaks chiral symmetry, however, $O_{LL}$
may mix under renormalisation with four wrong chirality operators, which we
denote $O_{\rm mix,i}$ for $i=1,\cdots,4$. Leading order chiral perturbation
theory predicts that
\be 
\langle \bar{K^0}| O_{LL} |K^0\rangle \propto M_K^2
\ee 
and, unfortunately, that 
\be 
\langle \bar{K^0}| O_{\rm mix,i} |K^0\rangle \propto 1 \, .
\ee 
This means that as the chiral limit is approached, these wrong chirality
operators will dominate the calculation.  However, for practical lattice
calculations we do not work in the chiral limit and, in fact, we argue that
for this calculation the mixing with such operators may be entirely neglected.
This may be seen by studying the pattern of chiral symmetry breaking using the
spurion field technique introduced in \cite{Blum:2001sr}; a simple application
of this method leads to the conclusion that the wrong chirality operators are
suppressed by a factor of $O((am_{\rm res})^{2})\approx 10^{-6}$.  A numerical
study, in the quenched approximation, to bound the of size of the
contributions of such operators has also been made. As reported on elsewhere
\cite{cd:jun}, these effects were, indeed, found to be negligible.

\begin{figure}[!ht]
\vspace{-1.0cm}
\begin{center}
\resizebox{7.2cm}{!}{\rotatebox{0}{\includegraphics{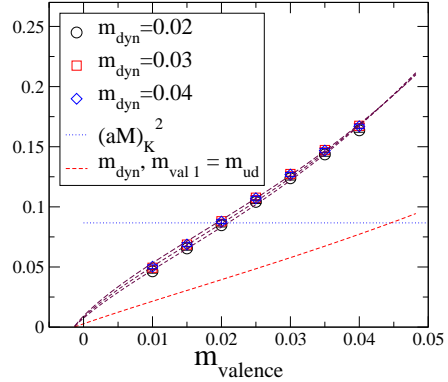}}}
\end{center}
\vspace{-1.8cm}
\caption{$(am_\pi)^2$ versus bare valance quark mass for all ensembles,
  together with the results of a fit to NLO partially-quenched chiral
  perturbation theory.}
\label{mass}
\end{figure}
\begin{figure}[!ht]
\vspace{-1.7cm}
\begin{center}
\resizebox{7.2cm}{!}{\rotatebox{0}{\includegraphics{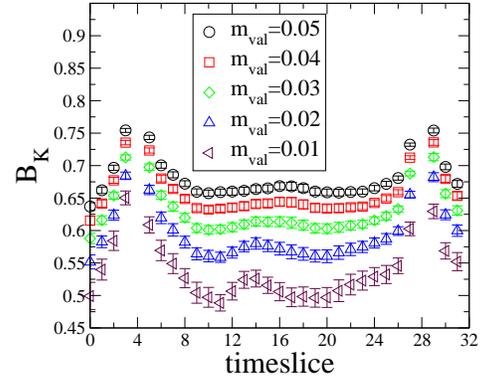}}}
\end{center}
\vspace{-1.8cm}
\caption{Bare $B_K$ versus timeslice for the $m_{\rm dyn}=0.04$ ensemble.}
\label{plat}
\vspace{-0.5cm}
\end{figure}

To calculate $B_K$ both the strange quark mass, $m_s$, and the average of the
up and down quark masses, $\overline{m}$, must be known. We extract these by
fitting the quark mass dependence of the pseudo-scalar and vector mesons to
the predictions of NLO partially-quenched chiral perturbation theory
(pq-$\chi_{\rm pt}$) and a linear ansatz, respectively. Requiring that the
pseudo-scalar and vector mesons have masses equal to the experimental value
for the pion and rho in the limit
\be 
m_{\rm dyn} = m_{\rm valence} =
\overline{m} \, , 
\ee
and the pseudo-scalar meson mass be eqaul to the kaon mass in the limit
\be 
m_{\rm dyn} = m_{\rm valence,1} = \overline{m} \ ; \ m_{\rm valence,2} = m_s
\, , 
\label{fit:ndeg}
\ee
gives a set of equations that we iteratively solve to find
$a\overline{m}=0.00017(11)$, $am_{s}=0.0446(29)$ and the lattice spacing
previously quoted. (Note these are bare quark masses; the renormalised quark
mass is defined as $Z_m(m+m_{res})$, where $Z_m$ is a scheme and scale
dependent renormalisation factor.) Figure \ref{mass} shows the square of the
pseudo-scalar meson versus (degenerate) valence quark mass, together with the
results of the fit to NLO pq-$\chi_{\rm pt}$. For the calculation of $B_K$
using only degenerate valence quarks, we will also make use of the strange
quark mass fixed by comparing the pseudo-scalar mass to the kaon mass in the
limit
\be
m_{\rm dyn} = \overline{m} \ ; \ m_{\rm valence} =
m_s/2 \, ,
\label{fit:deg}
\ee 
which leads to a value which is $\approx 10\%$ lower than the previous one.

We extract the bare value of $B_K$ from the ratio given in Eq.~\ref{eq:bk}
using (Coulomb) gauge-fixed wall-sources for the kaons at time-slices 4 and
28, varying the position of the operator. As in
\cite{Blum:2000kn,Blum:2001xb}, a combination of propagators with periodic and
anti-periodic boundary conditions are used to avoid contamination due to the
periodicity of the lattice. Figure \ref{plat} shows this ratio for the the
$m_{\rm dyn}=0.04$ ensemble for various degenerate valence quark masses. We
take our final values from an error-weighted average of the points between
time-slices 14 and 17, but see little change in our answer if this range is
varied.

\begin{figure}[!t]
\vspace{-1cm}
\begin{center}
\resizebox{7.2cm}{!}{\rotatebox{0}{\includegraphics{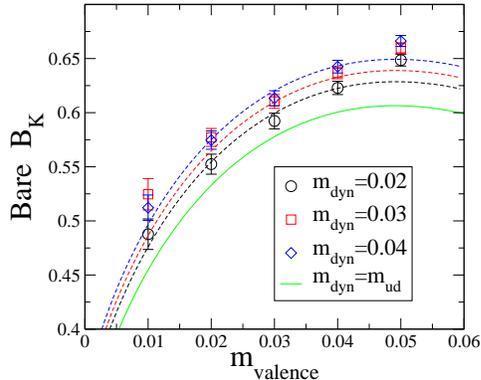}}}
\end{center}
\vspace{-1.8cm}
\caption{Bare, degenerate, $B_K$ versus valence quark mass, together with the
  results of the fit to NLO pq-$\chi_{\rm pt}$.}
\label{deg}
\vspace{-0.5cm}
\end{figure}

To extrapolate/interpolate to the physical point we again use the results of a
fit to pq-$\chi_{\rm pt}$, restricting the bare quark masses use to be $\le
0.04$ in an attempt to stay in the region of validity of NLO pq-$\chi_{\rm
pt}$.  Also, we do not include the $m_{\rm valence}=0.01$ data in our analysis
due to the bad quality of the plateau. Figure \ref{deg} shows the data for
degenerate valence quark masses together with the NLO pq-$\chi_{\rm pt}$
fit. Completely taking the limit given in Eq.\ref{fit:deg} leads to a final
result, when considering only degenerate valence quark masses, of \be B_K^{\rm
deg} = 0.509(18) \ee while including non-degenerate valence quark masses, and
taking the limit given in Eq.\ref{fit:ndeg}, gives \be B_K = 0.495(18) \, .
\ee While these value differ by less than the quoted error, as these errors
are correlated, this $3\%$ effect is \emph{statistically} well resolved.  We
quote both of these numbers in the $\overline{MS}$ scheme at $2{\rm GeV}$; the
needed combination of renormalisation factors to convert from the bare value
of $B_K$, $Z_{B_K}=Z_{LL}/Z_A^2$ \cite{cd:taku03}, is calculated using the NPR
technique of the Rome-Southampton group.

\section{CONCLUSIONS}

As part of the first large-scale project using dynamical domain wall fermions,
we have calculated the kaon B-parameter in two-flavour QCD. We present not
only a value using degenerate valence quark masses, of $B_K=0.509(18)$
($\overline{MS},2{\rm GeV}$), but also a value using non-degenerate masses
($m_s \neq m_u = m_d$), of $B_K=0.495(18)$ ($\overline{MS},2{\rm GeV}$). While
this difference is within the individual statistical errors, due to
correlations between the quantities, we find this difference to be
statistically well resolved.  These values are significantly below those found
in the quenched approximation. We caution that this work is based at
a single lattice spacing, on a single volume, and uses dynamical quark masses
which are large compared to the physical up and down quark masses together with
a quenched strange quark.

\bibliography{bibfile}
\bibliographystyle{h-elsevier}
\end{document}